\documentclass{elsart}
\usepackage{graphicx}
\usepackage{bm}
\usepackage{color}
\usepackage{amssymb}
\usepackage{amsmath}
\usepackage{url}

\newcommand \be{\begin{equation}}
\newcommand \ba{\begin{eqnarray}}
\newcommand \ee{\end{equation}}
\newcommand \ea{\end{eqnarray}}

\begin{document}

\begin{frontmatter}

\title{Is There a Real-Estate Bubble in the US?}

\author[ecust]{\small{Wei-Xing Zhou}},
\author[UCLA,nice]{\small{Didier Sornette}\thanksref{ADDR}}
\address[ecust]{State Key Laboratory of Chemical Reaction
Engineering, East China University of Science and Technology,
Shanghai 200237, China}
\address[UCLA]{Institute of Geophysics and Planetary Physics and
Department of Earth and Space Sciences, University of California,
Los Angeles, CA 90095}
\address[nice]{Laboratoire de Physique de la Mati\`ere Condens\'ee,
CNRS UMR 6622 and Universit\'e de Nice-Sophia Antipolis, 06108 Nice
Cedex 2, France}
\thanks[ADDR]{Corresponding author. Department of Earth and Space
Sciences and Institute of Geophysics and Planetary Physics,
University of California, Los Angeles, CA 90095-1567, USA. Tel:
+1-310-825-2863; Fax: +1-310-206-3051. {\it E-mail address:}\/
sornette@moho.ess.ucla.edu (D. Sornette)\\
\url{http://www.ess.ucla.edu/faculty/sornette/}}

\begin{abstract}
We analyze the quarterly average sale prices of new houses sold
in the USA as a whole, in the northeast, midwest, south,
and west of the USA, in each of the 50 states and the District of
Columbia of the USA, to determine whether they have grown
faster-than-exponential which we take as the diagnostic of a bubble.
We find that 22 states (mostly Northeast and West) exhibit clear-cut
signatures of a fast growing bubble. From the analysis
of the S\&P 500 Home Index, we conclude that the turning point of
the bubble will probably occur around mid-2006.
\end{abstract}

\begin{keyword}
Econophysics; Real estate; Bubble; Prediction
\PACS 89.65.Gh, 89.90.+n
\end{keyword}

\end{frontmatter}

\section{Is there a real-estate bubble in the US? Lessons from the
past UK bubble}
\label{sec:intro}

In the aftermath of the burst of the ``new economy'' bubble in 2000,
the Federal Reserve aggressively reduced short-term rate yields in
less than two years from 6$^{1/2}$~\% to 1~\% in June 2003 in an
attempt to coax forth a stronger recovery of the US economy. In
March 2003, we released a paper \footnote{See, W.-X. Zhou and D.
Sornette, \url{http://arXiv.org/abs/physics/0303028}} published a
few months later \cite{Zhousor1realestate} addressing the growing
apprehension at the time (see for instance \cite{housecards}) that
this loosening of the US monetary policy could lead to a new bubble
in real estate, as strong housing demand was being fueled by
historically low mortgage rates. As of March 2003, we concluded
that, ``while there is undoubtedly a strong growth rate, there is no
evidence of a super-exponential growth in the latest six years,''
giving `` no evidence whatsoever of a bubble in the US real estate
market'' \cite{Zhousor1realestate}.

More than two years have passed. During that period, the
historically low Fed rate of 1\% remained stable from June 2003 to
June 2004. Since June 2004, the Fed (specifically, the Federal Open
Market Committee (FOMC)) has increased its discount rate by
increments of 0.25\% at each of its successive meetings (the FOMC
holds 8 meetings per year): at the time of writing (end of May
2005), the last 0.25\% increase occurred on May 3rd, 2005 to yield a
short-term rate of 3\%, the next meeting of the FOMC being scheduled
on 29/30 June 2005 \footnote{Federal Open Market Committee,
$http://www.federalreserve.gov/FOMC/\#calendars$}. While the
short-term interest rates are following a steady upward trend of 2\%
per year since June 2004, long-term rates have not followed, some
going down while other long-term rates increasing only slightly.
Thus, long-term mortgage interest rates have remained extremely low
by historical standard. The Office of Federal Housing Enterprise
Oversight (OFHEO), the government unit tasked with regulating Fannie
Mae and Freddie Mac \footnote{Fannie Mae (resp. Freddie Mac) is a
stockholder-owned corporation chartered by the Federal Government in
1938 (resp. by Congress in 1970) to keep money flowing to mortgage
lenders in support of homeownership and rental housing.}, recently
published a research paper \footnote{Office of Federal Housing
Enterprise Oversight (OFHEO), Mortgage markets and the enterprises
(October 2004), prepared by V.L. Smith and L.R. Bowes
(\url{http://www.ofheo.gov/Research.asp})} stating that ``The
housing market achieved record levels of activity and contributed
significantly to the economic recovery ... Falling mortgage rates
stimulated housing starts and sales, and many refinancing borrowers
took out loans that were larger than those they paid off, providing
additional funds for consumption expenditures... According to
Freddie Mac, homeowners who refinanced in 2003 converted almost
\$139 billion in home equity into cash, up from \$105 billion in
2002.'' This has led to renewed worries that a real-estate bubble is
on its way.

The purpose of this paper is to revisit this question, using the more
than two additional
years of data since our previous analysis \cite{Zhousor1realestate}.
As explained in our previous paper on real-estate bubbles
\cite{Zhousor1realestate},
our analysis relies on a general theory of financial
crashes and of stock market instabilities developed in a series of
works (see \cite{SJB96,CriCrash99,JSL99,CriCrash00,SorJoh01,bookcrash} and
references therein). The main ingredient of the theory is the
existence of positive feedbacks in stock markets as well as in the
economy. Positive feedbacks, i.e., self-reinforcement, refer to the
fact that, conditioned on the observation that the market has
recently moved up (respectively down), this makes it more probable
to keep it moving up (respectively down), so that a large cumulative
move may ensue. The concept of ``positive feedbacks'' has a long
history in economics. It can occur for instance in the form of ``increasing
returns''-- which says that goods become cheaper the more of them
are produced (and the closely related idea that some products, like
fax machines, become more useful the more people use them).
Positive feedbacks, when unchecked, can produce runaways until the
deviation from equilibrium is so large that other effects can be
abruptly triggered and lead to rupture or crashes. Alternatively, it
can give prolonged depressive bearish markets.
There are many mechanisms leading to positive feedbacks including
investors' over-confidence, imitative behavior and herding between investors,
refinancing releasing new cash re-invested in houses, lower
requirement margins due to uprising prices, and so on.
Such positive feedbacks provide the fuel for the development of
speculative bubbles, by
the mechanism of cooperativity, that is, the interactions and
imitation between investors may lead to collective behaviors similar
to crowd phenomena. Different types of collective regimes are
separated by so-called critical points which, in physics, are widely
considered to be one of the most interesting properties of complex
systems. A system goes critical when local influences propagate over
long distances and the average state of the system becomes
exquisitely sensitive to a small perturbation, {\it i.e.} different
parts of the system become highly correlated. Another characteristic
is that critical systems are self-similar across scales: at the
critical point, an ocean of traders who are mostly bearish may have
within it several continents of traders who are mostly bullish, each
of which in turns surrounds seas of bearish traders with islands of
bullish traders; the progression continues all the way down to the
smallest possible scale: a single trader \cite{Wilson}. Intuitively
speaking, critical self-similarity is why local imitation cascades
through the scales into global coordination. Critical points are
described in mathematical parlance as singularities associated with
bifurcation and catastrophe theory. At critical points, scale
invariance holds and its signature is the power law behavior of
observables.

Mathematically, these ideas are captured by the power law
\begin{equation}
    \ln[p(t)] = A + B (t_c-t)^m~,
    \label{Eq:PL}
\end{equation}
where $p(t)$ is the house price or index, $t_c$ is an estimate of
the end of a bubble so that $t<t_c$ and $A,B,m$ are coefficients. If
the exponent $m$ is negative, $\ln[p(t)]$ is singular when $t \to
t_c^-$ and $B>0$ ensuring that $\ln[p(t)]$ increases. If $0<m<1$,
$\ln[p(t)]$ is finite but its first derivative $d\ln[p(t)]/dt$ is
singular at $t_c$ and $B<0$ ensuring that $\ln[p(t)]$ increases.
Extension of this power law (\ref{Eq:PL}) takes the form of
log-periodic power law (LPPL) for the logarithm of the price
\begin{equation}
\ln[p(t)] = A + B(t_c -t)^{m} + C(t_c-t)^{m}
\cos\left[\omega\log(t_c-t)-\phi\right], \label{Eq:lnpt}
\end{equation}
where $\phi$ is a phase constant and $\omega$ is the angular
log-frequency. This first version (\ref{Eq:lnpt}) amounts to assume
that the potential correction or crash at the end of the bubble is
proportional to the total price \cite{CriCrash99}. In contrast, a
second version assumes that the potential correction or crash at the
end of the bubble is proportional to the bubble part of the total
price, that is to the total price minus the fundamental price
\cite{CriCrash99}. This gives the following price evolution:
\begin{equation}
p(t) = A + B(t_c -t)^{m} + C(t_c-t)^{m}
\cos\left[\omega\log(t_c-t)-\phi\right]~. \label{Eq:pt}
\end{equation}
As explained in \cite{Zhousor1realestate,bookcrash}, we diagnose a
bubble using these models by demonstrating a faster-than-exponential
increase of $p(t)$, possibly decorated by log-periodic oscillations.

Before presenting the result of our analysis using these models on
the US real-estate bubble, it is appropriate to discuss how our
detection of a bubble in the UK real-estate market fared since March
2003. In \cite{Zhousor1realestate}, we reported ``unmistakable
signatures (log-periodicity and power law super-exponential
acceleration) of a strong unsustainable bubble'' for the UK
real-estate market. We identified two potential turning points in
the UK bubble reported in Tables 2 and 3 of
\cite{Zhousor1realestate}: end of 2003 and mid-2004. The former
(resp. later) was based on the use of formula (\ref{Eq:lnpt}) (resp.
(\ref{Eq:pt})). These predictions were performed in Feb. 2003 (again
our paper was released in early March 2003 on an electronic archive
\footnote{W.-X. Zhou and D. Sornette,
\url{http://arXiv.org/abs/physics/0303028}}). We stress that these
turning points can be either crashes or changes of regimes according
to the theory coupling rational expectation bubbles with collective
herding behavior described in
\cite{CriCrash00,JSL99,predictds,JS05as}. In other words, the theory
describes bubbles and their end but not the crash itself: the end of
a bubble is the most probable time for a crash, but a crash can
occur earlier (with low probability) or not at all; the possibility
that no crash occurs is necessary for the bubble to exist,
otherwise, rational investors would anticipate the crash and, by
backward reasoning, would make it impossible to develop.

\begin{figure}
\begin{center}
\includegraphics[width=14cm]{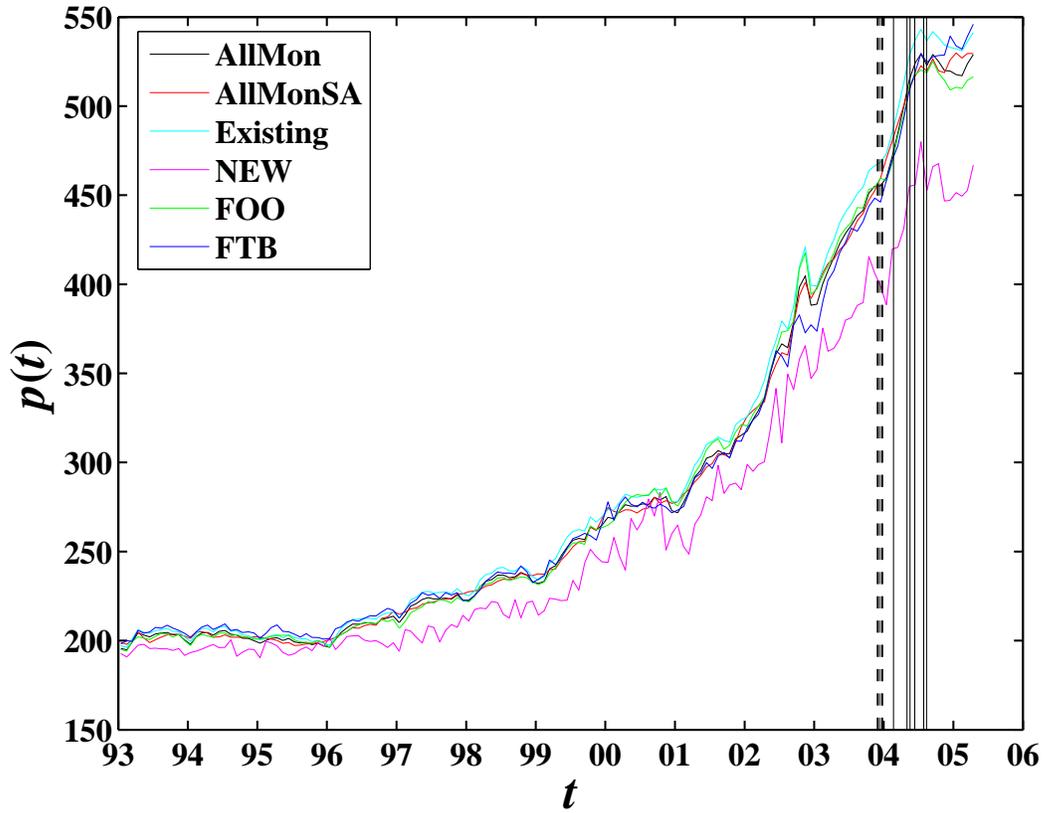}
\end{center}
\caption{Plot of the UK Halifax house price indices from 1993 to
April 2005 (the latest available quote at the time of writing). The
two groups of vertical lines correspond to the two predicted turning
points reported in Tables 2 and 3 of \cite{Zhousor1realestate}: end
of 2003 and mid-2004. The former (resp. later) was based on the use
of formula (\ref{Eq:lnpt}) (resp. (\ref{Eq:pt})). These predictions
were performed in Feb, 2003.} \label{FigUK}
\end{figure}

Figure \ref{FigUK} plots the UK Halifax house price index (HPI)
\footnote{The Halifax house price index has been used extensively by
government departments, the media and businesses as an authoritative
indicator of house price movements in the United Kingdom. This index
is based on the largest sample of housing data and provides the
longest unbroken series of any similar UK index. The monthly house
price index data are retrieved from the web site of HBOS
\url{http://www.hbosplc.com/view/housepriceindex/housepriceindex.asp.}.
The six time series are the following. AllMon: All Houses (All
Buyers); AllMonSA: All Houses (All Buyers) (seasonally adjusted);
Existing: Existing Houses (All Buyers); New: New Houses (All
Buyers); FOO: Former Owner Occupiers (All Houses); FTB: First Time
Buyers (All Houses).} from 1993 to April 2005 (the latest available
quote at the time of writing). The two groups of vertical lines
correspond to the two predicted turning points mentioned above. The
first set of predicted turning points (dashed lines in Figure
\ref{FigUK}) anticipated by half-a-year the turning point which
occurred mid-2004 as predicted by the second set.

Our analysis presented below uses three data sets: (1) the regional
data (Northeast, Midwest, west, south and USA as a whole) of the
quarterly average sale prices of new houses up to the fourth
quarter of 2004 (the latest data available); (2) the house price
index of individual states (50 states $+$ DC), up to Q1 of 2005,
quarterly data; and (3) daily data of the S\&P 500 Home Index, up to
May 6, 2005. We first present in section \ref{sec:Power-law} a
broad-brush analysis using the exponential versus power law models
of house price appreciation for the whole continental US and then by
regions. We then turn to a state-by-state analysis which leads to a
partition into three classes: (i) non-bubbling states, (ii)
recent-bubbling states and (iii) clearly-bubbling states. For the
states for which a bubble seems to be clearly established according
to our criterion, we provide a first estimation of the critical time
of the end of the bubble. We then turn to the more elaborate LPPL
models (\ref{Eq:lnpt}) and (\ref{Eq:pt}) and a nonlinear extension,
using the daily data of the S\&P 500 Home Index up to May 6, 2005.

\section{Evidence of a US real-estate bubble by faster-than-exponential growth}
\label{sec:Power-law}

\begin{figure}
\begin{center}
\includegraphics[width=14cm]{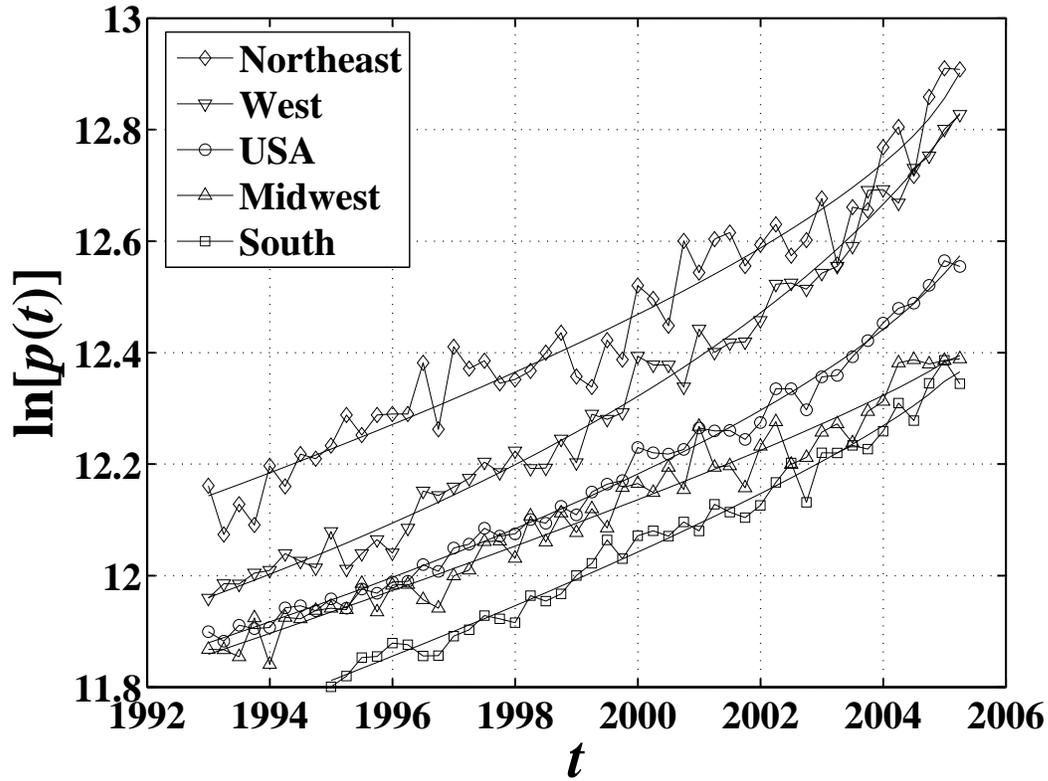}
\end{center}
\caption{Quarterly average sale prices of new houses sold in all the
states of the USA and by regions (Northeast, Midwest, South and
West) from 1993 to the first quarter of 2005 as a function of time
$t$. The smooth lines are the power-law fits (\ref{Eq:PL}) to the
five data sets. The values of fitted parameters for the West are
$t_c=2008.3$, $m=0.18$, $A=15.48$, and $B=-2.16$ with the r.m.s. of
the fit residuals being 0.0275. The values of the fitted parameters
for USA are $t_c=2006.0$, $m= 0.55$, $A=12.75$, and $B= -0.212$ with
the r.m.s. of the fit residuals being 0.0174. The values of fitted
parameters for the Northeast are $t_c=2005.2$, $m= 0.64$, $A=12.92$,
and $B= -0.154$ with the r.m.s. of the fit residuals being 0.0485.
The values of fitted parameters for the South are $t_c=2005.2$, $m=
0.78$, $A=12.37$, and $B= -0.092$ with the r.m.s. of the fit
residuals being 0.0215. The values of fitted parameters for the
Midwest are $t_c=2005.2$, $m= 0.84$, $A=12.40$, and $B= -0.066$ with
the r.m.s. of the fit residuals being 0.0308.} \label{Fig:Regional}
\end{figure}

Figures \ref{Fig:Regional} show the quarterly average sale prices of
new houses sold in all the states of the USA as well as in the four
main regions, Northeast, Midwest, South and West, from 1993 to the
fourth quarter of 2004 as a function of time $t$. The smooth curve
is the power-law fit (\ref{Eq:PL}) to the data. Except for the
midwest and south regions, one can observe a strong upward curvature
in these linear-logarithmic plots, which characterize a
faster-than-exponential price growth (recall that an exponential
growth would qualify as a straight line in such linear-logarithmic
plots). The existence of a strong upward curvature characterizing a
faster-than-exponential growth is quantified by the relatively small
values of the exponent $m$ ($=0.55$ for all states, $=0.64$ for the
Northeast region, $=0.18$ for the West region).

\begin{figure}
\begin{center}
\includegraphics[width=14cm]{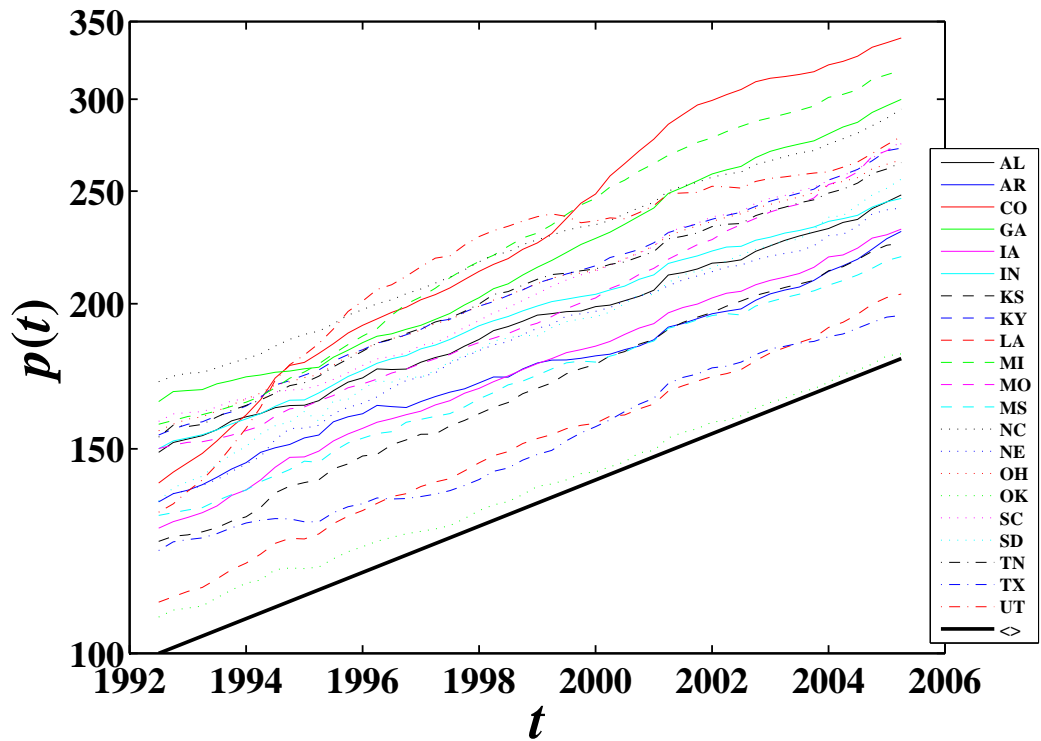}
\end{center}
\caption{(Color online) Quarterly HPI in the 21 states which have an
approximately constant exponential growth, qualified by a linear
trend in a linear-logarithmic scale. The thick straight line at the
bottom of the figure is the average over all 21 states corresponding
to an annual growth rate of 4.6\% over the last 13 years. The
corresponding states are given in the legend. Note that Colorado
seems to be on a faster trend. } \label{Fig:51STS:Exp}
\end{figure}

\begin{figure}
\begin{center}
\includegraphics[width=14cm]{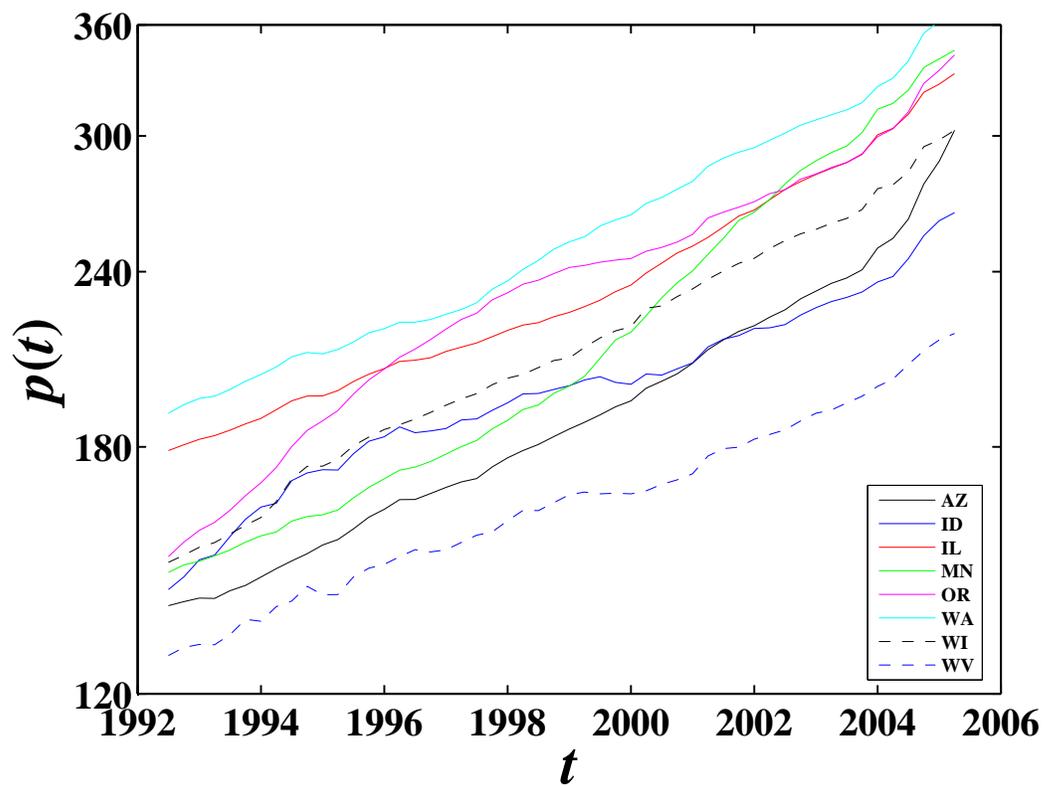}
\end{center}
\caption{(Color online) Quarterly HPI in the 8 states exhibiting a
recent upward acceleration following an approximately constant
exponential growth rate. The corresponding states are given in the
legend.} \label{Fig:51STS:ExpAc}
\end{figure}

\begin{figure}
\begin{center}
\includegraphics[width=14cm]{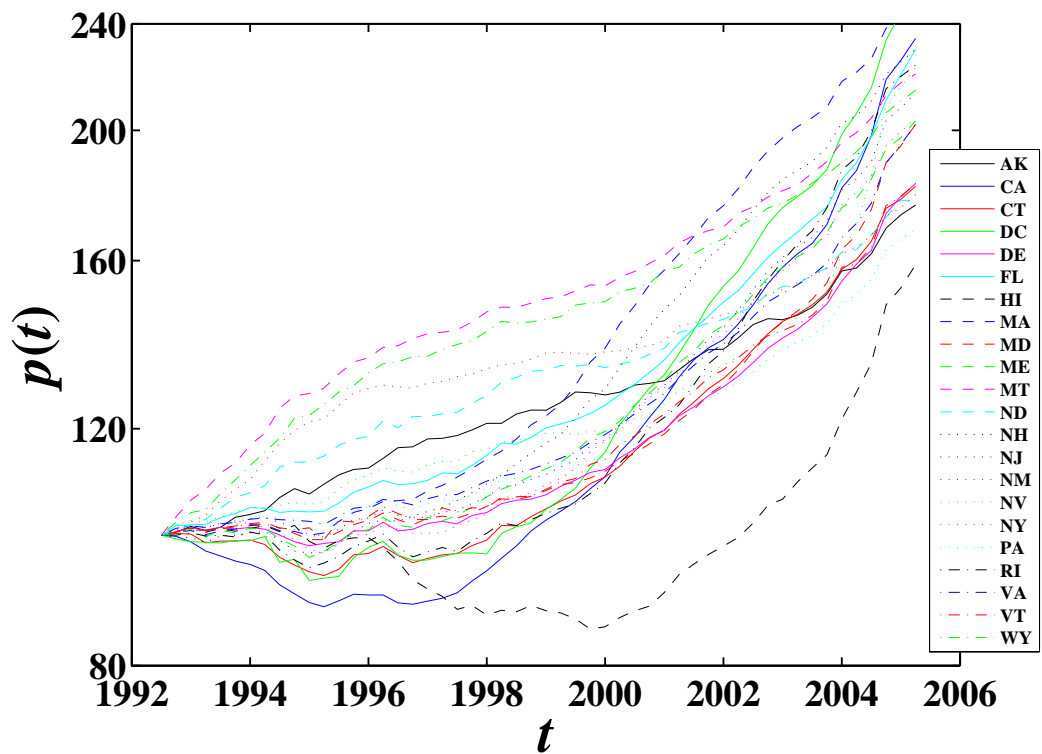}
\end{center}
\caption{(Color online) Quarterly average HPI in the 21 states and
in the District of Columbia (DC) exhibiting a clear upward
faster-than-exponential growth. For better representation, we have
normalized the house price indices for the second quarter of 1992 to
100 in all 22 cases. The corresponding states are given in the
legend.} \label{Fig:51STS:PL}
\end{figure}

\begin{figure}
\begin{center}
\includegraphics[width=14cm]{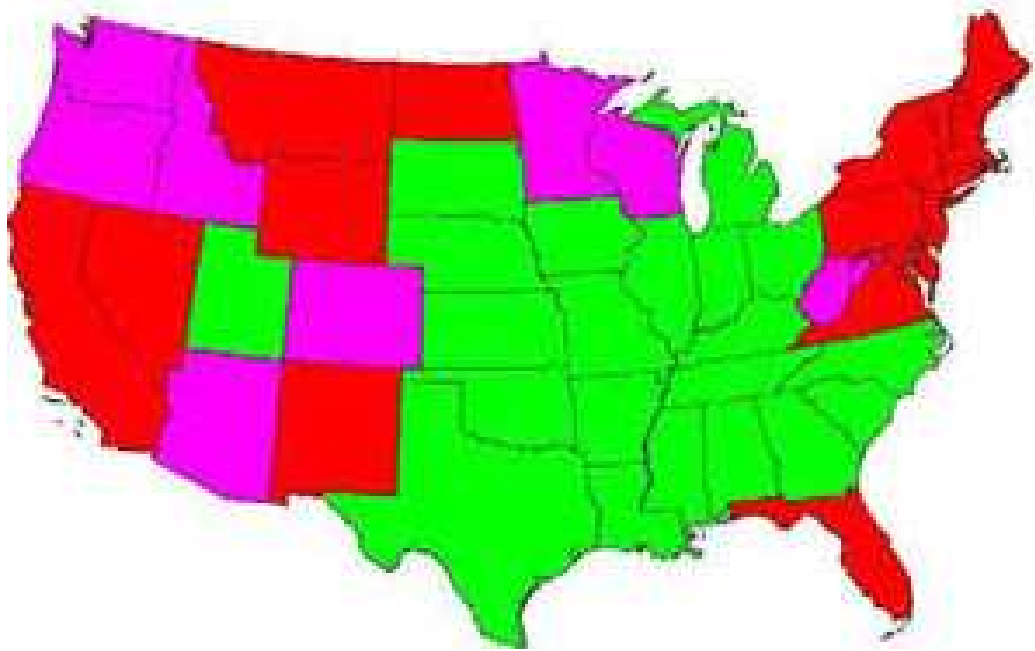}
\end{center}
\caption{Geographical synopsis of the classification in three families:
the first family of non-bubbling states
of figure \ref{Fig:51STS:Exp} in green, the
second family of recently-bubbling states in figure
\ref{Fig:51STS:ExpAc} in magenta and
the third family of clearly bubbling states in figure
\ref{Fig:51STS:PL} in red.
Hawaii and Alaska, both in red, are not drawn here.} \label{Fig:USA:Map}
\end{figure}

To have a closer look, we examined quarterly data of House Price
Index (HPI) for each individual state. Rather than following a
formal procedure and developing sophisticated statistical tests, the
obvious differences between the price trajectories in the different
states led us to prefer a more intuitive approach consisting of
classifying the different states according to how strongly they
depart from a steady exponential growth. We found three families,
shown in figures \ref{Fig:51STS:Exp}, \ref{Fig:51STS:ExpAc} and
\ref{Fig:51STS:PL}. Figure \ref{Fig:51STS:Exp} shows the quarterly
HPI in the 21 states which have an approximately constant
exponential growth, qualified by a linear trend in a
linear-logarithmic scale. The thick straight line at the bottom of
the figure is the average over all 21 states corresponding to an
annual growth rate of 4.6\% over the last 13 years (we did not use
the data prior to 1992 to avoid contamination by the turning point
of the previous bubble in 1991). Figure \ref{Fig:51STS:ExpAc} shows
the quarterly HPI in the 8 states exhibiting a recent upward
acceleration following an approximately constant exponential growth
rate. Figure \ref{Fig:51STS:PL} shows the quarterly HPI in the 22
states exhibiting a clear upward faster-than-exponential growth.
These 22 states thus exhibit the hallmark of a real-estate bubble.
Figure \ref{Fig:USA:Map} provides a geographical synopsis of this
classification in three families: the first family of figure
\ref{Fig:51STS:Exp} is green, the second family of figure
\ref{Fig:51STS:ExpAc} is magenta, and the third family of figure
\ref{Fig:51STS:PL} is red. As often discussed by commentators,
prices have accelerated mostly to the Northeast and West regions,
which is consistent with Fig.~\ref{Fig:Regional}.

\begin{figure}
\begin{center}
\includegraphics[width=14cm]{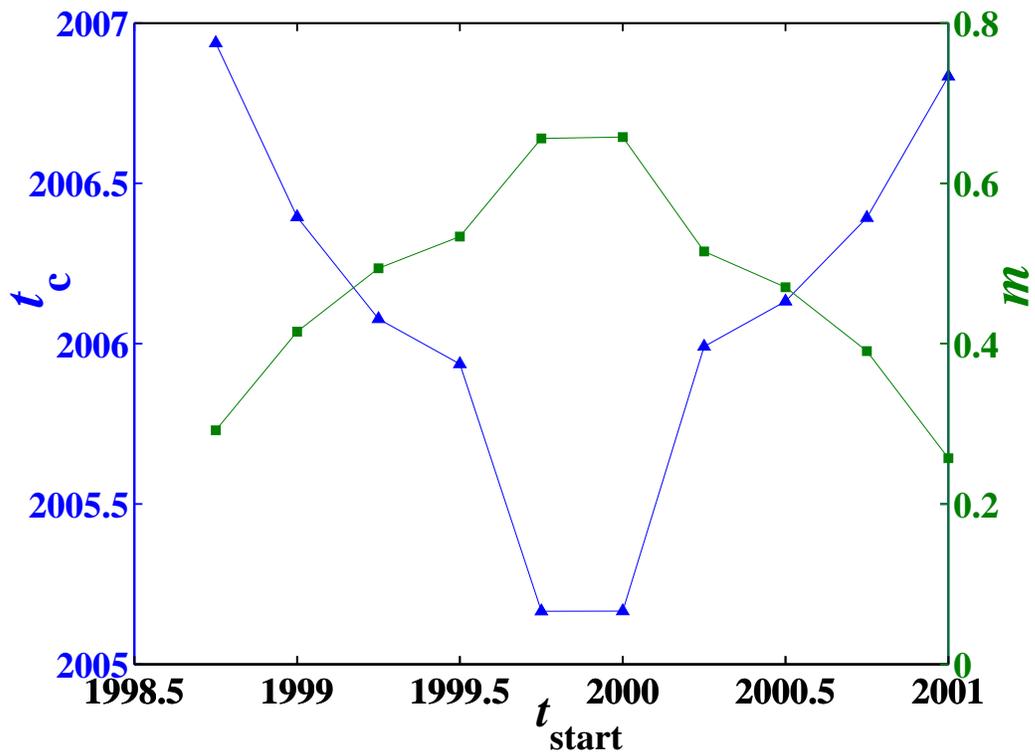}
\end{center}
\caption{Critical time $t_c$ and
exponent $m$ as a function of $t_{\rm start}$, obtained by fitting
with the power law (\ref{Eq:PL}) (with the modification that
$t_c-t$ is replaced by $|t_c-t|$ to allow for a more robust estimation)
the price trajectory obtained
by averaging over the 22 states of figure \ref{Fig:51STS:PL}.
} \label{Fig51STS_PL_AveFit}
\end{figure}

Consider the third family of 22 states where we diagnose a bubble,
as shown in figure \ref{Fig:51STS:PL}. Can a power law fit with
(\ref{Eq:PL}) reveal the end of the bubble? Such turning point is in
principle measured by the time $t_c$ in expression (\ref{Eq:PL}),
which gives the time at which the bubble should end. In order to get
less noisy data, we averaged over the 22 price trajectories of
figure \ref{Fig:51STS:PL} and then fitted the obtained average with
(\ref{Eq:PL}) (with the modification that $t_c-t$ is replaced by
$|t_c-t|$ to allow for a more robust estimation) over a time
interval from $t_{\rm start}$ to the last available data point
(2005Q1). Figure \ref{Fig51STS_PL_AveFit} shows the obtained
critical time $t_c$ and exponent $m$ as a function of $t_{\rm
start}$. Varying $t_{\rm start}$ allows us to test for sensitivity
with respect to the different time periods and assess the robustness
of the results. Not surprisingly, we find that the fitted $t_c$ is
close to the last data points for some $t_{\rm start}$, a result
which has been found to
systematically characterize power law behaviors \cite{bookcrash}.
Therefore, the power law fit is not very reliable to determine the end of the
real-estate bubble. However, the relative stability of $m$ in the
range $0.1-0.5$ as
a function of $t_{\rm start}$, which characterizes a faster-than-exponential
growth, indicates that the simple power law (\ref{Eq:PL}) is already
a good model.

\section{Extension to the LPPL model and discussion}

The previous tests performed in
\cite{SJB96,CriCrash99,JSL99,CriCrash00,SorJoh01,bookcrash} (and
references therein) show that the problem with the too-large
sensitivity of $t_c$ in the simple power law model with respect to
the few last data points are alleviated by using the more
sophisticated LPPL models (\ref{Eq:lnpt}) and (\ref{Eq:pt}). Here,
we use both LPPL models (\ref{Eq:lnpt}) and (\ref{Eq:pt}) as well as
the so-called 2nd-order Landau LPPL introduced in \cite{SJ97}
\footnote{See also
$http://www.ess.ucla.edu/faculty/sornette/prediction/index.asp\#prediction$
for a recent application to the US stock market}. In a nutshell, the
2nd-order Landau LPPL extends the LPPL model by allowing for a first
nonlinear correction which amounts to combining two log-frequencies
$\omega_0$ (close to $t_c$) and $\omega_\infty$ (far from $t_c$).

\begin{figure}
\begin{center}
\includegraphics[width=14cm]{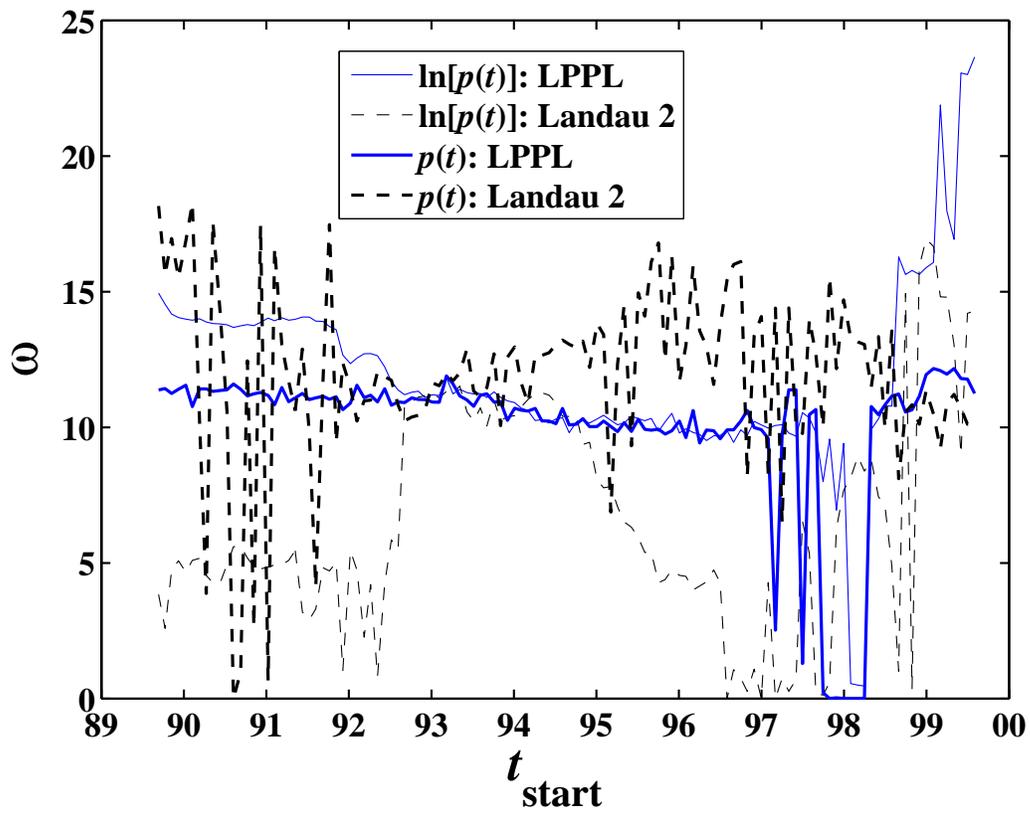}
\end{center}
\caption{Dependence of $\omega$ for the LPPL model and of $\omega_0$
for the 2nd-order Landau LPPL model as a function of $t_{\rm start}$
obtained by fitting the S\&P 500 Home Index up to May 6, 2005.}
\label{FigSP5Home_omega}
\end{figure}

\begin{figure}
\begin{center}
\includegraphics[width=14cm]{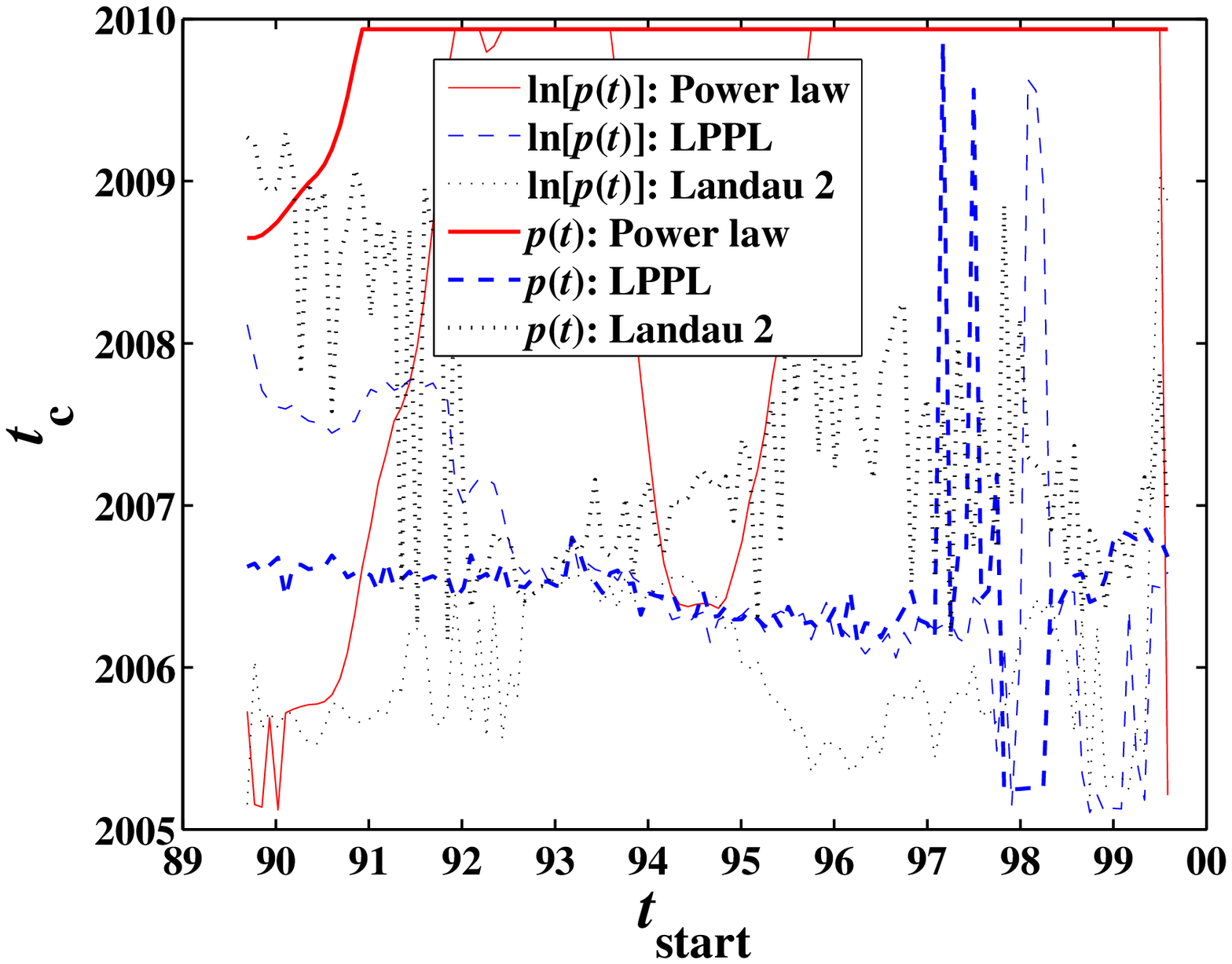}
\end{center}
\caption{Predicted critical time $t_c$ as a function of $t_{\rm
start}$ obtained from the fits with the LPPL and the 2nd-order
Landau LPPL models as in figure \ref{FigSP5Home_omega}. }
\label{FigSP5Home_tc}
\end{figure}

We fit the daily data of the S\&P 500 Home Index to the LPPL and
2nd-order Landau LPPL models in a time interval from $t_{\rm start}$
to the last available data point (April 2005). Figure
\ref{FigSP5Home_omega} presents the dependence of $\omega$ for the
LPPL model and of $\omega_0$ for the 2nd-order Landau LPPL model as
a function of $t_{\rm start}$. We find $\omega \approx 10$ and
$\omega_0$ oscillating between $\omega \approx 10$ and $\omega/2$,
as we expect from the generic existence of harmonics (see our
previous extended discussions in \cite{SZpr3mdas,ZSkskaa}). The
stability of $\omega_0$ and the compatibility between the two
descriptions is a signature of the robustness of the signal. Figure
\ref{FigSP5Home_tc} shows the predicted critical time $t_c$ as a
function of $t_{\rm start}$ obtained from the fits with the LPPL and
the 2nd-order Landau LPPL models. The large spreads of values for
$t_{\rm start}$ earlier than 1993 reflects the fact that the bubble
has really started only after 1993.

We observe a good stability of the
predicted $t_c \approx$ mid-2006 for the two LPPL models
(\ref{Eq:lnpt}) and (\ref{Eq:pt}).
The spread of $t_c$ is larger for the second-order LPPL fits but
brackets mid-2006.
As mentioned before, the power law fits are not reliable.
We conclude that the turning point of the bubble will probably occur
around mid-2006.


\newpage

\clearpage

\clearpage

\clearpage

\clearpage

\clearpage

\clearpage

\clearpage

\clearpage

\clearpage


\begin{thebibliography}{}


\bibitem{housecards} House of cards, The Economist, May 29, 2003.

\bibitem{CriCrash00} A. Johansen, O. Ledoit and D. Sornette,
Crashes as critical points, International Journal of Theoretical and
Applied Finance 3 (2), 219-255 (2000).

\bibitem{CriCrash99} A. Johansen and D. Sornette, Critical crashes,
Risk, Vol 12, No. 1, p.91-94 (1999).

\bibitem{JS05as} A. Johansen and D. Sornette,
Endogenous versus exogenous crashes in financial markets,
in press in ``Contemporary Issues in International Finance''
(Nova Science Publishers, 2005)
(http://arXiv.org/abs/cond-mat/0210509)

\bibitem{JSL99} A. Johansen, D. Sornette and O. Ledoit,
Predicting financial crashes using discrete scale invariance,
Journal of Risk 1 (4), 5-32 (1999)

\bibitem{bookcrash} D. Sornette, Why Stock Markets Crash
(Critical Events in Complex Financial Systems), Princeton University
Press, Princeton, NJ, 2002.

\bibitem{SJ97} D. Sornette and A. Johansen, Large financial
crashes, Physica A 245, N3-4, 411-422 (1997).

\bibitem{SorJoh01} D. Sornette and A. Johansen, Significance of
log-periodic precursors to financial crashes, Quantitative Finance
1, 452-471 (2001).

\bibitem{SJB96} D. Sornette, A. Johansen and J.-P. Bouchaud,
Stock market crashes, precursors and replicas, J. Phys. I France 6,
167-175 (1996).

\bibitem{SZpr3mdas} D. Sornette and W.-X. Zhou,
    The US 2000-2002 market descent: how much longer and deeper?
Quantitative Finance 2, 468-481 (2002).

\bibitem{predictds} D. Sornette and W.-X. Zhou,
Predictability of large future changes in major financial indices,
in press in the International Journal of Forecasting (2005)
(http://arXiv.org/abs/cond-mat/0304601)

\bibitem{Wilson} K. G. Wilson,
Problems in Physics with many scales of length, Scientific American
241, 158-179 (1979).

\bibitem{Zhousor1realestate}
W.-X. Zhou and D. Sornette, 2000-2003 real estate bubble in the UK
but not in the USA, Physica A 329, 249-263 (2003).

\bibitem{ZSkskaa} W.-X. Zhou and D. Sornette,
Renormalization group analysis of the 2000-2002 anti-bubble in the US
S\&P 500 index: Explanation of the hierarchy of 5 crashes and prediction,
Physica A 330, 584-604 (2003).

\end{thebibliography}
\end{document}